# Unusual Circumstances of the 2024 June 8 GLE

Nat Gopalswamy*[(1)], Pertti Makela[(2)], Hong Xie[(2)], Sachiko Akiyama[(2)], Seiji Yashiro[(2)], and Atul Mohan[(2)]
(1) NASA Goddard Space Flight Center, Greenbelt, MD, 20771, USA, https://cdaw.gsfc.nasa.gov
(2) The Catholic University of America, Washington DC 20064, USA

## Abstract

Ground Level Enhancement (GLE) in large solar energetic particle (SEP) events is indicative of protons accelerated to GeV energies. Almost all GLE events are associated with sustained gamma-ray emission (SGRE) from the Sun because the latter require >300 MeV protons that are readily present during GLEs. Here we report on the 2024 June 8 GLE event, which has the distinction of not being associated with an SGRE event. All the associated phenomena typical of SGRE events were present: a fast and wide CME, a major solar flare, and an intense type II radio bursts that extend from the metric to kilometric wavelength domains. There was a data gap of ~51 min, but the SGRE is expected to last for hours. We suggest the east-west asymmetry in the flow of energetic particles from the shock is likely to be the reason for the lack of SGRE emission.

## 1. Introduction

Ground level enhancement (GLE) in solar energetic particle (SEP) events indicate the acceleration of protons to GeV energies [1]. GLEs are indicative of high-energy (>100 MeV) sustained gamma-ray emission (SGRE) from the Sun because occurrence of GeV protons guarantees the presence of >300 MeV protons responsible for SGRE [2]. Coronal mass ejections (CMEs) underlying GLEs and SGREs are among the fastest and associated type II bursts live the longest [3]. In fact, all GLEs are associated with SGRE when there are overlapping observations since the 1980s. The Large Area Telescope (LAT [4]) onboard the Fermi mission started observing SGREs in solar cycle 24. There were only two GLEs in this cycle (2012 May 17 and 2017 September 10) and both were associated with SGREs. Five GLEs occurred in solar cycle 25. GLEs 73 (2021 October 28) and 77 (2025 November 11) were associated with SGRE. GLE 74 (2024 May 11) occurred during a major Fermi/LAT data gap, so no high-energy gammas were observed. GLEs 75 (2024 June 08) and GLE 76 (2024 November 21) were not associated with SGRE. GLE 76 originated from a behind the west limb (S10W118) eruption, so high-energy protons might not have precipitated to the front side. Lack of SGRE during GLE75 is quite puzzling and hence motivated this study.

## 2. Observations

GLE 75 was a small event, with an estimated peak increase of ~3% above the background. Detailed properties of the GLE has been reported in [5], so we give a brief summary here. Figure 1 shows the GLE count rate as a function of time. The GLE event started, peaked, and ended at 01:55, 02:40, and 04:10 UT, respectively. The eruption occurred in NOAA active region (AR) 13697 located at ~S17W60. The eruption involved an M9.7 flare immediately preceded by an M3.3 flare. Both flares were associated with coronal mass ejections (CMEs) with possible CME interaction.

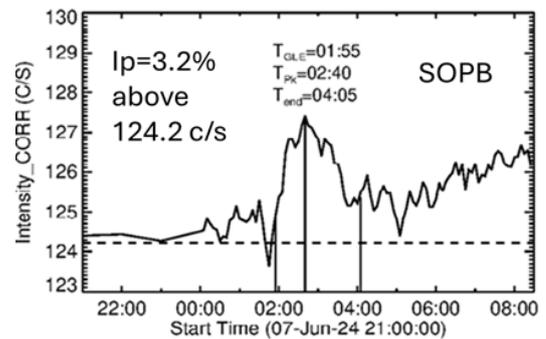

**Figure 1.** Intensity of the 2024 June 8 GLE expresses as counts per second (c/s) as observed by the South Pole Bare (SOPB) Neutron Monitor. The background count rate is 124.2 c/s, so the peak percentage enhancement above the background is 3.2%. Data obtained from the Oulu GLE database (https://gle.oulu.fi).

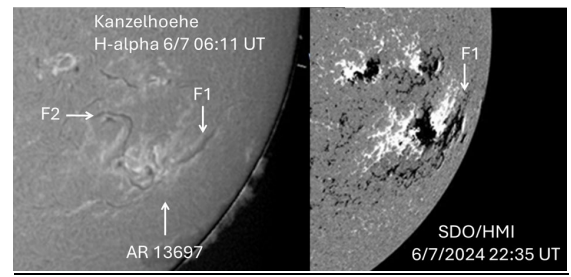

**Figure 2.** (left) H-alpha filaments F1 (eruptive) and F2 (stable) in AR 13697. (right) SDO/HMI magnetogram taken a couple of hours before the eruption shows the AR with multiple neutral lines; F1's neutral line is pointed to.

### 2.1 A Complex Eruption

AR 13697 was quite extended from W40 to W75 in longitude and S15 to S30 in latitude. The magnetic structure was complex with multiple neutral lines in the magnetogram obtained by the Solar Dynamics

Observatory's Helioseismic and Magnetic Imager (SDO/HMI). An H-alpha image taken on June 7 by the Kanzelhoehe Solar Observatory shows two prominent filaments one in the eastern part of the AR and the other from the western part (see Fig. 2). A narrow filament is seen near the sunspots.

Figure 3 shows the evolution of the M3.3 flare in the SDO's Atmospheric Imaging Assembly (AIA) 1600 Å images. The eruption was triggered by an emerging flux region at the south end of filament F1 as seen in the image at 00:37 UT. The filament seems to have broken off near the emerging flux site with one fragment going eastward and another going northward.

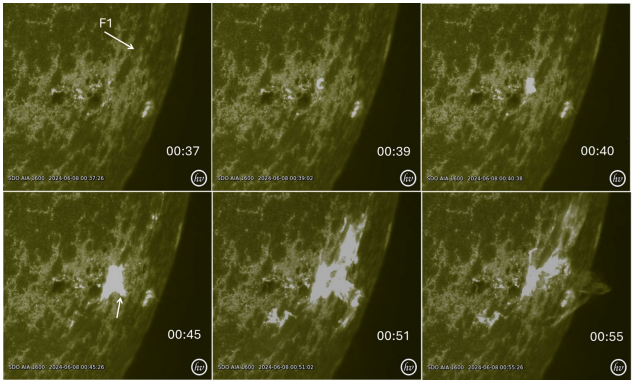

Figure 3. Triggering of the eruption of filament F1due to emerging flux at its southern end (upper middle image at 00:39 UT). The filament broke off and moved east (arrow in the 00:45 UT image) and north. The filament fragment is seen above the limb at 00:55 UT. The M3.3 flare started, peaked, and ended at 00:39, 00:51, and 00:57 UT.

Figure 4 shows the M9.7 eruption starting with two compact brightenings near the ribbons of the previous M3.3 flare. From this region, another filament fragment erupted, which was very fast (see Fig. 5). The flare evolution involved brightening of the ribbons and extending generally to the north. The ribbons are located to the north-eastern edge of the active region, making it closer to the limb (~W70) about 10° away from the AR location.

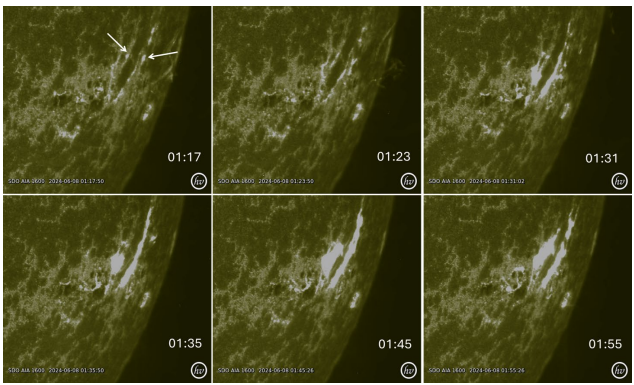

**Figure 4.** Evolution of the M9.7 flare as seen in SDO/AIA 1600 Å images. The ribbon locations are the same as those of the first flare. The M9.7 flare started, peaked, and ended at 01:23 UT, 01:49 UT and 02:19 UT, respectively.

Figure 5 shows the three filament fragments involved in this eruption. Fragments labelled 1 and 2 started during the M3.3 flare, while 3 started during the M9.7 flare. Fragment 3 was relatively fast and became the core of the fast CME. Coronagraph images show that there was a faint CME overlying filament fragments 1 and 2 that was overtaken by the second CME reported in the next subsection.

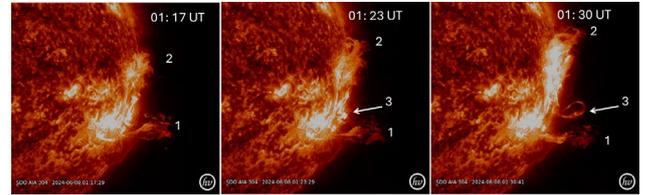

**Figure 5.** Filament fragments during the 2024 June 8 M9.7 eruption. Fragments 1 and 2 started in association with the M3.3 flare, while 3 started as part of the M9.7 flare.

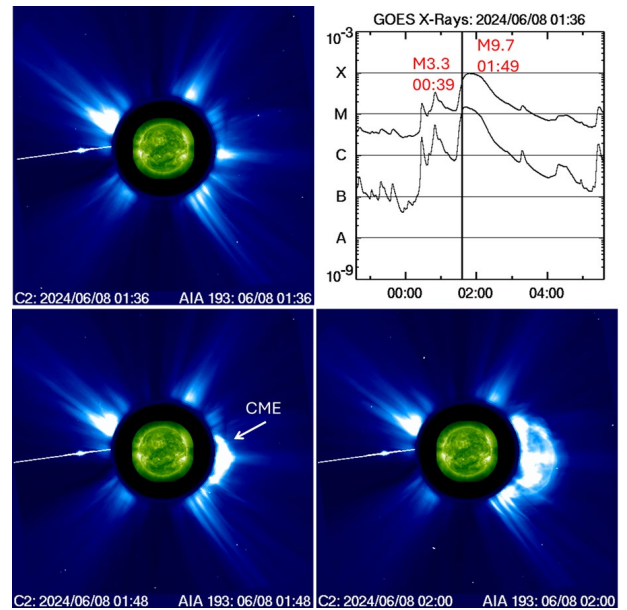

**Figure 6**. (top) SOHO/LASCO C2 image before the CME appeared and the GOES light curve. (bottom) CME at first appearance (01:48 UT) and at 02:00 UT.

## 2.2 A Fast CME

Figure 6 shows the fast CME observed by the Large Angle and Spectrometric Coronagraph (LASCO) onboard the Solar and Heliospheric Observatory (SOHO) using its C2 telescope. The CME started as a normal west limb CME, but expanded rapidly to become a full halo CME. The CME first appeared in LASCO/C2 FOV at 01:48 UT with leading edge (LE) at 3.28 Rs, which moved out to 4.73 Rs, at 02:00 UT. The sky-plane speed between these heights was ~1408 km/s. The last two height-time data points (04:30 UT, 23.36 Rs) and (04:42 UT, 24.88 Rs) in LASCO/C3 FOV yield a sky-plane speed of 1476 km/s indicating continued acceleration in the coronagraph FOV. The early phase of the CME was observed by the Extreme Ultraviolet Imager (EUVI), and inner coronagraph COR1 onboard the Solar Terrestrial Relations Observatory (STEREO) Ahead spacecraft. The height-time data points (01:30:04 UT, 1.23 Rs), (01:32:30 UT, 1.35 Rs) from EUVI, and (01:35:00 UT, 1.54 Rs) from COR1 give a speed of 699 km/s. Given the STEREO-Ahead location at W15, the solar source was at W55, so the speed deprojects to 853 km/s. Taken with the deprojected speeds in LASCO/C2 (1498 km/s) and C3 (1570 km/s) FOVs, we confirm the continued acceleration. The acceleration in the coronagraph FOV is typically much smaller than the initial acceleration obtained from the CME speed and flare rise time. Using 1518 km/s for CME speed and flare rise time = 26 min, we get an initial acceleration of ~1 km s$^{-2}$. This is at the lower end of values for GLE CMEs. There are only two other GLEs with initial acceleration smaller than 1 km s$^{-2}$ [6].

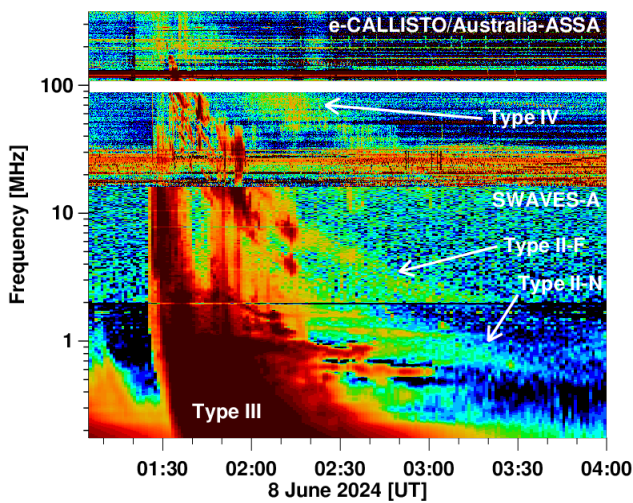

**Figure 7.** Australia ASSA CALLISTO dynamic spectrum combined with that from STEREO/WAVES.

## 2.3 A Type II Radio Burst

Figure 7 shows a composite dynamic spectrum from the e-CALLISTO telescope ASSA in Australia, and the radio and plasma wave (WAVES) instrument onboard STEREO. The eruption involved a type III burst group, a type II burst, and a type IV burst. The type II activity started at 01:27 UT and lasted until 01:34 UT in the metric domain and continued into the decameter-hectometric (DH) domain. At 01:31 UT, the type II burst started at a high frequency of ~220 MHz (fundamental). The harmonic component can also be discerned in the dynamic spectrum. There were a few pairs with fundamental harmonic structure at frequencies below 100 MHz. Such a complex type II spectrum is typical of GLE events because the shock is fast and radio emission is produced throughout the shock surface passing through varying density structures. Type II bursts in the metric domain continue into the DH domain marked as Type II -F (F for flank). Distinct from Type II-F is type II-N (N for nose) as marked in Fig. 8. Type II-N starts around 6 MHz at ~01:40 UT. Note that the nose and flank components are not harmonically related. The dynamic spectrum also includes a type IV burst and barely enters the DH domain. There was also type III-like bursts emanating from the type II burst during 01:44 to 02:06 UT. The type II burst continued to drift to lower frequencies and ended only around 10:30 UT with an additional component during the last three hours (not shown). The type II burst lasted for ~9 hours with an ending frequency of ~250 kHz.

The shock formation height was obtained by extrapolating the early height-time measurements to the time of metric type II onset (01:31 UT) as 1.27 Rs (1.55 Rs deprojected). At the time type II-N started, the CME LE height reached 2.46 Rs.

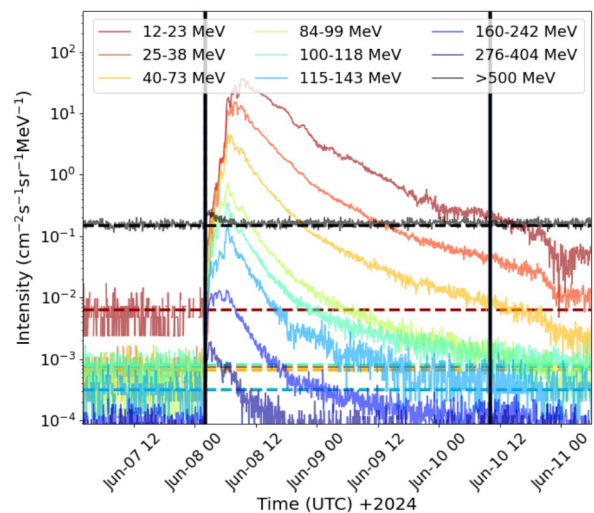

**Figure 8.** SEP intensity profiles from GOES-16 at various energy channels from 12 MeV to >500 MeV. The highest energy channel is typically close to the GLE time profile. The horizontal dashed lines show the channel backgrounds.

## 2.4 An SEP Event

The eruption was accompanied by a large SEP event with a proton intensity of ~1000 pfu in the >10 MeV GOES energy channel. The >100 MeV integral channel also showed a high intensity of ~10 pfu [7]. Figure 8 shows SEP intensities in several differential energy channels including the >500 MeV integral channel. The >500 MeV protons lasted for several hours until ~15 UT. The 276-404 MeV channel shows particles until the end of June 9. The intense SEP event is consistent with the strong metric-to-kilometric type II burst indicative of an energetic CME-driven shock. While the CME speed is close to the average speed of SEP-producing CMEs, the high intensity is consistent with interacting CMEs [8].

The first arriving GeV particles are only ~3 min behind electromagnetic emissions, so the solar particle release (SPR) time can be estimated as 01:52 UT (3 min before Earth arrival). The CME height at SPR is computed as 4.0 Rs by extrapolating the measurement at 01:48 UT to 01:52

UT and deprojecting. This means the CME/shock had to travel from 1.55 Rs to 4 Rs before SPR happens. The SPR ($h_{SPR}$ in Rs) is consistent with the parabolic relation with source longitude ($\lambda$ in degrees): $h_{SPR} = 2.55 + [(\lambda - 51)/35]^2$. For $\lambda = 70°$, we expect $h_{SPR} = 2.84$ Rs as compared to the observed 4 Rs. The agreement is reasonable and within the scatter in the original plot [9].

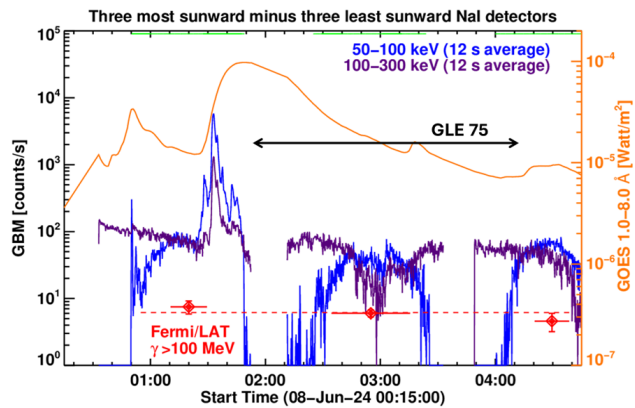

**Figure 9.** A summary of the flare in soft X-rays and hard X-rays (Fermi/GBM). The Fermi/LAT data points (red symbols) do not show an SGRE event. The horizontal bars on the red points indicate the overall exposure time in each orbit. The GLE duration is shown by the double headed arrow. The green bars at the top denote Fermi daytime. There is a small data gap (~51 min) after GOES peak.

## 2.5 Fermi Gamma-ray Observations

Under the shock paradigm for SGRE, the copious production of >300 MeV protons (Fig. 8) should indicate an SGRE event. Figure 9 summarizes X-ray and gamma-ray emissions associated with GLE 75. The GOES light curve shows the M3.3 and M9.7 flares. During the impulsive phase of the M9.7 flare, the Gamma-ray Burst Monitor (GBM) [10] on Fermi observed >100 keV hard X-rays, that lasted for ~ 15 min. Such long-duration HXR bursts have been found in all SGRE events when simultaneous observations are available [11]. Also shown are Fermi/LAT >100 MeV gamma-ray flux along with their background level. The three data points (at 01:20 UT, 02:55 UT, and 04:30 UT) obtained around the eruption time are at or below the background with no indication of an SGRE event. The 01:20 UT data point is just before the start of the M9.7 flare, while the 02:55 UT data point is after the end of the flare. The Sun exposure began at 02:40 UT and ended at 03:15 UT) Therefore, there is no information on the prompt (impulsive phase) gamma-ray event. SGRE typically starts around the peak of the GOES soft X-ray flare (01:49 UT here). There is no Fermi/LAT data during the 51 minutes from the flare peak to 02:40 UT. Given the long duration of type II burst, one would have expected an hours-long SGRE. The type II duration ($d_{II}$) is known to be linearly related to the gamma-ray duration ($d_g$) according to $d_g = 0.9 d_{II} - 0.8$ expressed in hours [3]. Since $d_{II} = 9$ h for our event, we expect $d_g = 7.3$ h. Thus, we conclude that it is likely that GLE 75 lacked SGRE.

One possible explanation is that the accelerated particles from the shock were not able to reach the photosphere. This can happen when the high-energy particles precipitate to a high magnetic field region and get mirrored back before reaching the photosphere. There is nothing out of the ordinary in the magnetic structure of the AR and surroundings, so we do not expect mirroring to affect all high energy particles. The second possibility is that the high-energy particles did precipitate to the photosphere, but not on the frontside. This can happen when high energy particles predominantly precipitate on the west side of the CME, which was behind the west limb. Such a possibility has been invoked to explain the predominance of behind the east the limb eruptions [12].

## 3. Discussion and Summary

We reported on GLE 75, which is a small event (~3% above background). The underlying CME was typical of SEP-producing CMEs (~1500 km/s). The CME initial acceleration (~1 km s$^{-2}$), shock formation height (1.55 Rs), and solar particle release height (4.0 Rs) are all typical of GLE events. The >10 MeV intensity is rather high (~1000 pfu) often observed in CMEs interacting with preceding ones (as is the case here). These properties are typically shared by SGRE events because of the overlap in the energy of the underlying particles. We attribute the lack of SGRE during GLE 75 due to possible flow of the high-energy particles from the shock nose predominantly to the western side of the CME with very little precipitating on the front side. GLEs with similar near-limb locations are known to be associated with SGRE, so additional factors such as CME flux rope geometry may also play a role.

## 4. Acknowledgements

We thank the SOHO, STEREO, SDO, Fermi, GOES, Oulu Cosmic-ray, Kanzelhoehe solar, and e-CALLISTO teams for making their data available online. Work supported by the STEREO project.